\documentclass{eptcs}
\providecommand{\event}{ADG 2021} 
\usepackage{amsmath}
\usepackage{amssymb}

\usepackage[T1]{fontenc}
\usepackage[english]{babel}
\usepackage{tikz,pgf}
\usepackage[labelfont=rm]{subcaption}
\usepackage{hyperref}

\tikzstyle{vertex}=[circle, draw, fill=black, minimum size=5.5pt, inner sep=0.8pt]
\newcommand{\vertex}{\node[vertex]}
\tikzstyle{edge}=[very thick]
\tikzstyle{edge1}=[edge,draw=red]

\newcommand{\N}{\mathbb{N}}
\newcommand{\Q}{\mathbb{Q}}
\newcommand{\R}{\mathbb{R}}
\newcommand{\C}{\mathbb{C}}

\newcommand{\edg}{\mcal{E}}
\newcommand{\Lam}{\operatorname{Lam}}
\newcommand{\lam}{\mathrm{Lam}}
\newcommand{\lamII}{\lam}
\newcommand{\Mtwo}{M}
\newcommand{\maxlamII}[1]{\Mtwo(#1)}
\newcommand{\maxlamIIn}{\maxlamII{n}}
\newcommand{\leftquot}[2]{{}^{#1}\!{#2}}
\newcommand{\rightquot}[2]{{#2}^{#1}}
\newcommand{\modop}{\,\mathrm{mod}\,}

\newcommand{\subtract}{\mathop{\backslash}}
\newcommand{\quotient}{\mathop{/}}

\newcommand{\vectwo}[2]{\binom{#1}{#2}}
\newcommand{\vecxy}{\vectwo{x}{y}}
\newcommand{\mcal}{\mathcal}

\newtheorem{definition}{Definition}
\newtheorem{theorem}{Theorem}
\newtheorem{proposition}{Proposition}

\def\orcidID#1{\unskip$^{\mbox{\href{https://orcid.org/#1}{\scriptsize{[#1]}} }}$}

\title{Realizations of Rigid Graphs\thanks{Supported by the Austrian Science 
Fund (FWF): F5011-N15. Based on joint work with Jose Capco, Matteo Gallet, 
Georg Grasegger, Niels Lubbes, Josef Schicho, and Elias Tsigaridas.}
}

\author{Christoph Koutschan\orcidID{0000-0003-1135-3082}
\institute{Johann Radon Institute for Computational and Applied Mathematics (RICAM) \\
4040 Linz, Austria}
\email{christoph.koutschan@ricam.oeaw.ac.at}
} 

\def\titlerunning{Realizations of Rigid Graphs}
\def\authorrunning{Koutschan}

\begin{document}
\maketitle
\begin{abstract}
  A minimally rigid graph, also called Laman graph, models a planar framework
  which is rigid for a general choice of distances between its vertices. In
  other words, there are finitely many ways, up to isometries, to realize such
  a graph in the plane. Using ideas from algebraic and tropical geometry, we
  derive a recursive formula for the number of such realizations. Combining
  computational results with the construction of new rigid graphs via gluing
  techniques, we can give a new lower bound on the maximal possible number of
  realizations for graphs with a given number of vertices.
\end{abstract}

\section{Introduction}

The theory of rigid graphs forms a fascinating research area in the
intersection of graph theory, computational algebraic geometry, and
algorithms. The study of rigid structures, also called frameworks, was
originally motivated by mechanics, and it goes back at least to the 19th
century.  Besides being a very interesting mathematical subject, rigid graphs
and the underlying theory of Euclidean distance geometry have meanwhile found
a large number of applications ranging from robotics and bioinformatics to
sensor network localization and architecture.

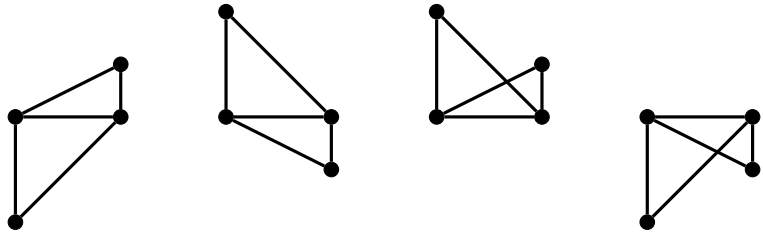
\begin{figure}[ht]
  \begin{center}
    \begin{tikzpicture}[scale=0.7]
      \begin{scope}
	\vertex (a1) at (0,-2) {};
	\vertex (b1) at (0,0) {};
	\vertex (c1) at (2,0) {};
	\vertex (d1) at (2,1) {};
	\path[edge] (a1)edge(b1) (b1)edge(c1) (c1)edge(a1) (b1)edge(d1) (c1)edge(d1);
      \end{scope}
      \begin{scope}[xshift=1cm]
	\vertex (a2) at (3,2) {};
	\vertex (b2) at (3,0) {};
	\vertex (c2) at (5,0) {};
	\vertex (d2) at (5,-1) {};
	\path[edge] (a2)edge(b2) (b2)edge(c2) (c2)edge(a2) (b2)edge(d2) (c2)edge(d2);
      \end{scope}
      \begin{scope}[xshift=2cm]
	\vertex (a3) at (6,2) {};
	\vertex (b3) at (6,0) {};
	\vertex (c3) at (8,0) {};
	\vertex (d3) at (8,1) {};
	\path[edge] (a3)edge(b3) (b3)edge(c3) (c3)edge(a3) (b3)edge(d3) (c3)edge(d3);
      \end{scope}
      \begin{scope}[xshift=3cm]
	\vertex (a4) at (9,-2) {};
	\vertex (b4) at (9,0) {};
	\vertex (c4) at (11,0) {};
	\vertex (d4) at (11,-1) {}; 
	\path[edge] (a4)edge(b4) (b4)edge(c4) (c4)edge(a4) (b4)edge(d4) (c4)edge(d4);
      \end{scope}
    \end{tikzpicture}
  \end{center}
  \caption{All four realizations of the minimally rigid graph with four vertices.}
  \label{figure:realizations}
\end{figure}

Suppose that we are given a graph $G$ with edge set~$E$.  We consider the set
of all possible realizations (embeddings) of the graph in the Euclidean plane
such that the lengths of the edges coincide with some prescribed edge labeling
$\lambda\colon E\rightarrow \R_{>0}$.  Edges and vertices are allowed to
overlap in such a realization.  For example, suppose that $G$ has four
vertices and is a complete graph minus one edge.
Figure~\ref{figure:realizations} shows all possible realizations of $G$ up to
rotations and translations, for a certain edge labeling.  We address the
following problem:
\begin{quote}
 \itshape{For a given graph determine its number of realizations for a general
   edge labeling, up to rotations and translations.}
\end{quote}
Here we say that a property holds for a \emph{general edge labeling} if it
holds for all edge labelings belonging to a dense open subset of the vector
space of all edge labelings.

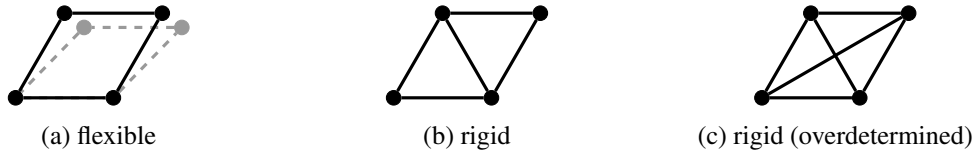
\begin{figure}
  \begin{center}
    \begin{subfigure}[b]{0.3\textwidth}
      \begin{center}
        \begin{tikzpicture}[scale=1.3]
          \vertex (a) at (0,0) {};
          \vertex (b) at (1,0) {};
          \vertex (c) at (0.5,0.866025) {};
          \vertex (d) at (1.5,0.866025) {};
          \begin{scope}
            \vertex[black!40!white] (c2) at (0.7,0.72) {};
            \vertex[black!40!white] (d2) at (1.7,0.72) {};
            \draw[edge,black!40!white,dashed] (a) -- (b) -- (d2) -- (c2) -- (a) -- cycle;
          \end{scope}
          \draw[edge] (a) -- (b) -- (d) -- (c) -- (a) -- cycle;
        \end{tikzpicture}
        \caption{flexible}
        \label{figure:starrheit:flexible}
      \end{center}
    \end{subfigure}
    \begin{subfigure}[b]{0.3\textwidth}
      \begin{center}
        \begin{tikzpicture}[scale=1.3]
          \vertex (a) at (0,0) {};
          \vertex (b) at (1,0) {};
          \vertex (c) at (0.5,0.866025) {};
          \vertex (d) at (1.5,0.866025) {};
          \draw[edge] (a) -- (b) -- (d) -- (c) -- (a) -- cycle;
          \draw[edge] (b) -- (c);
        \end{tikzpicture}
        \caption{rigid}
        \label{figure:starrheit:rigid}
      \end{center}
    \end{subfigure}
    \begin{subfigure}[b]{0.3\textwidth}
      \begin{center}
        \begin{tikzpicture}[scale=1.3]
          \vertex (a) at (0,0) {};
          \vertex (b) at (1,0) {};
          \vertex (c) at (0.5,0.866025) {};
          \vertex (d) at (1.5,0.866025) {};
          \draw[edge] (a) -- (b) -- (d) -- (c) -- (a) -- cycle;
          \draw[edge] (b) -- (c);
          \draw[edge] (a) -- (d);
        \end{tikzpicture}
        \caption{rigid (overdetermined)}
        \label{figure:starrheit:overdetermined}
      \end{center}
    \end{subfigure}
  \end{center}
  \caption{Graphs and their state of rigidity}
  \label{figure:starrheit}
\end{figure}

The realizations of a graph can be considered as physical structures in the
plane, which consist of rods that are connected by rotational joints.  If a
graph together with an edge labeling admits infinitely (resp.\ finitely) many
realizations up to rotations and translations, then the corresponding planar
structure is flexible (resp.\ rigid), see Figure~\ref{figure:starrheit}.

A graph is called \emph{generically rigid} (or isostatic) if a general edge
labeling yields a rigid realization. No edge in a generically rigid graph can
be removed without losing rigidity, that is why such graphs are also called
\emph{minimally rigid} in the literature.  The complete graph on four vertices
$K_4$ is for instance not considered to be minimally generically rigid, since for a
general choice of edge lengths it will not have a realization: imagine you
would have to add the missing edge in either of the realizations depicted in
Figure~\ref{figure:realizations} --- for most prescribed lengths of the new
edge this will not be possible, see also Figure~\ref{figure:starrheit}.
Hilda Pollaczek-Geiringer~\cite{Geiringer1927} and independently
Gerard Laman~\cite{Laman1970} characterized the property of generic rigidity
in terms of the number of edges and vertices of the graph and its subgraphs,
hence such objects are also known as \emph{Laman graphs}.

\begin{theorem}
  \label{theorem:laman_graph}
  A graph $G=(V,E)$ is minimally (generically) rigid if and only if
  $\vert E\vert = 2\vert V\vert-3$, and for every subgraph $G'=(V',E')$
  with at least two vertices it holds $\vert E' \vert \leq 2\vert V' \vert-3$.
\end{theorem}

All finitely many realizations of a Laman graph can be obtained as the
solution set of a system of quadratic polynomial equations, where the edge
labels are either given by concrete numbers or interpreted as parameters.  In
general it is difficult to produce results on the number of real solutions of
such systems.  In such situations, one often switches to a complex setting;
this also enables us to apply results from algebraic geometry.  Hence, from
now on, we consider edge labelings with complex numbers, and we are interested
in the number of \emph{complex solutions}, up to an equivalence relation on
$\C^2$ generalizing the direct isometries of $\R^2$; this number is the same
for any general edge labeling, so we call it the \emph{Laman number} of the
graph~$G$, denoted by $\Lam(G)$.  For some graphs up to $8$ vertices, this
number had been computed using random values for the edge
labels~\cite{JacksonOwen2012} --- this means that it is very likely, but not
absolutely certain, that these computations give the true numbers. Upper and
lower bounds for Laman numbers are considered in
\cite{Emiris2013,Steffens2010,Borcea2004}.  Note that for many Laman graphs
there exists a (real) edge labeling such that the number of real realizations
equals precisely the Laman number.  However, there are graphs for which the
Laman number gives only an upper bound on the number of real realizations.

Our main result is a combinatorial algorithm that computes the number of
complex realizations of any given Laman graph; it is much more efficient than
just solving the corresponding nonlinear system of equations.  The algorithm
and its correctness proof are presented in detail in~\cite{SymbolicGroup}.
Using a supercomputer, we apply this algorithm to a large collection of Laman
graphs and identify among all Laman graphs with $n$ vertices ($n\leq12$) the
one with the maximal Laman number. This allows us to derive better lower
bounds on the number of realizations~\cite{Lowerbounds}. In the following, we
provide a concise summary of these results, focusing on the main ideas and the
algorithmic point of view.

\section{Computing the Laman Number}\label{sec:laman_number}

We write $G = (V,E)$ to denote a finite graph~$G$ with vertices~$V$ and
edges~$E$. An edge~$e$ between two vertices $u$ and $v$ is denoted by
$\{u,v\}$; this notation expresses the fact that all graphs considered here
are \emph{undirected}.

Using nonnegative real labels for the edge lengths, the number of realizations
in~$\R^2$ for a general edge labeling is not well-defined, since it heavily
depends on the actual labeling and not only on the graph. For example, the
complete graph $K_3$ permits two different realizations (one being the
reflection of the other) for almost all edge labelings that satisfy the
triangle inequality, while it has none for all other labelings.  In order to
define a number that depends only on the graph, we switch to a complex
setting. In order to keep notations simple, we take the convention that the
edge labelings give the squared distances between vertices.
\begin{definition}
  \label{definition:realization}
  Let $G = (V,E)$ be a graph.
  \begin{itemize}
  \item A \emph{labeling} of $G$ is a function $\lambda \colon E \longrightarrow \C$.
    The pair $(G, \lambda)$ is called a \emph{labeled graph}.
  \item A \emph{realization} of $G$ is a function
    $\rho\colon V \longrightarrow\C^2$. 
    Let $\lambda$ be a labeling of $G$: we say that a realization $\rho$ is 
    \emph{compatible with} $\lambda$ if for each $e \in E$ the distance
    between its endpoints agrees with its label:
    \[
      \lambda(e) \, = \, \bigl\langle \rho(u)-\rho(v), \rho(u)-\rho(v) \bigr\rangle,\quad
      e = \{u,v\},
    \]
    where $\left\langle x, y \right\rangle = x_1y_1 + x_2y_2$.
  \end{itemize}	
  A labeled graph $(G, \lambda)$ is \emph{realizable} if and only if there 
  exists a realization $\rho$ that is compatible with the edge labeling~$\lambda$.
	
  We say that two realizations of a graph $G$ are \emph{equivalent} if and
  only if there exists a direct isometry $\sigma$ of $\C^2$ between them,
  where $\sigma$ is a map of the form
  \begin{align*}
    \vecxy  &\longmapsto A\cdot \vecxy +b,
  \end{align*}
  where $A\in\C^{2\times 2}$ is an orthogonal matrix with determinant $1$ and $b\in\C^2$.
\end{definition}

\begin{definition}
	\label{definition:rigid_graph}
	A labeled graph $(G, \lambda)$ is called \emph{rigid} if it
	is realizable and	there are only finitely many realizations compatible with $\lambda$, up to 	equivalence.
\end{definition}

Our main interest is to count the number of realizations of generically rigid
graphs, namely graphs for which almost all realizable labelings induce rigidity.

\begin{definition}
	\label{definition:generically_rigid}
	A graph $G$ is called \emph{generically realizable} if for a general 
	labeling $\lambda$ the labeled graph $(G, \lambda)$ is realizable. A graph $G$ 
	is called \emph{generically rigid} if for a general labeling $\lambda$ the 
	labeled graph $(G, \lambda)$ is rigid.
\end{definition}

\noindent
The number of realizations can be found by solving the following system of equations
\begin{equation*}
  \Bigl( (x_u - x_v)^2 + (y_u - y_v)^2 \, = \, \lambda_{uv} \Bigr)_{\{u,v\} \in E}\,.
\end{equation*}
Equivalently, we can study the map $r_G$ whose preimages of
$(\lambda_{uv})_{\{u,v\}\in E}$ correspond to the solutions of the above
system
\begin{equation*}
  r_G\colon \C^{2|V|} \longrightarrow \C^{E}, \quad
  (x_v, y_v)_{v \in V} \; \longmapsto \; \Bigl( (x_u - x_v)^2 + (y_u - y_v)^2 \Bigr)_{\{u,v\} \in E}\,.
\end{equation*}
Still we get infinitely many solutions due to translations and rotations.
Translations can be eliminated by moving one vertex to the origin.
In order to handle rotations
we perform the following transformation
$x_v+i y_v \rightarrow x_v$,\;
$x_v-i y_v \rightarrow y_v$.
Then the above equations become
\begin{align*}
   \Bigl( (x_u - x_v)(y_u - y_v) \, = \, \lambda_{uv} \Bigr)_{\{u,v\} \in E}\,.
\end{align*}
In this way, solutions that differ only by a rotation are the same in a suitable projective setting.
If we transform the map $r_G$ accordingly we obtain a map whose degree is finite and gives the sought number of realizations.

In order to set up a recursive formula for the degree of the map, we want to be able to handle
the two factors $(x_u - x_v)$ and $(y_u - y_v)$ independently.
To do this we duplicate the graph, and for technical reasons
we allow a more general class of graphs.
The resulting concept is roughly speaking a pair of graphs $(G,H)$ with a bijection between their sets of edges.
We identify edges by this bijection.
\begin{definition}
  \label{definition:bigraph}
  A \emph{bigraph} is a pair of undirected graphs $(G, H)$ --- allowing
  several components, multiple edges and self-loops --- where $G = (V, \edg)$
  and $H = (W, \edg)$.  The set $\edg$ is called the set of \emph{biedges},
  and there are two maps that assign to each $e\in\edg$ the corresponding
  vertices in $V$ and $W$, respectively.  Note that $G$ and $H$ are in general
  different graphs but there is a bijection between their sets of edges.
\end{definition}

We define the Laman number $\Lam(B)$ of a bigraph $B$ as the degree of an
associated map defined in a similar way as $r_G$.  Moreover, we show that the
Laman number of a graph equals the Laman number of the corresponding bigraph.
\begin{proposition}\label{prop:laman_number}
  The number of realizations~$\Lam(G)$ of a Laman graph~$G$ is equal to the
  Laman number~$\Lam(B)$ of the bigraph $B=(G,G)$.
\end{proposition}

The idea for proving the recursion formula in
Theorem~\ref{theorem:laman_number} is inspired by tropical geometry: we
consider the equation system over the field of Puiseux series; an algebraic
relation between Puiseux series implies a piecewise linear relation between
their orders.  We encode these piecewise linear relations in a combinatorial
data which we call \emph{bidistance}.  A bidistance is a pair of functions
from the edges of a bigraph to the rational numbers~$\Q$, which satisfies
certain conditions.  Using a bidistance $d$ of a bigraph~$B$ we can define a
new bigraph $B_d$ with the same number of edges.  The solutions of the
equations for the bigraph $B$ that correspond to the bidistance~$d$ are in
bijection with the solutions of the equations for~$B_d$.  Then the solutions
for~$B$ are partitioned by the bidistances, implying the following formula for
the Laman number:
\begin{equation*}
  \Lam(B) \, = \, \sum_{d} \Lam(B_d).
\end{equation*}
From this we finally show the combinatorial recursion formula.  For doing so
we prove that $\Lam(B_d)$ is either easy to compute or the product of two
Laman numbers of bigraphs with fewer edges each.  We need some more notation
to state the theorem.

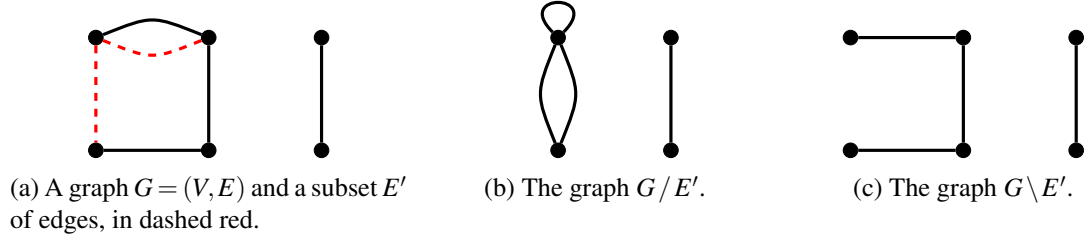
\begin{figure}
  \begin{center}
    \begin{subfigure}[t]{0.33\textwidth}
      \centering
      \begin{tikzpicture}[scale=1.5]
        \vertex (a) at (0,0) {};
        \vertex (b) at (1,0) {};
        \vertex (c) at (1,1) {};
        \vertex (d) at (0,1) {};
        \vertex (e) at (2,0) {};
        \vertex (f) at (2,1) {};
        
        \path[]
        (d) edge[edge1, dashed] (a)
        (a) edge[edge] (b)
        (b) edge[edge] (c)
        (e) edge[edge] (f)
        ;
        \draw[edge1, dashed] (d) .. controls (0.5, 0.8) .. (c);
        \draw[edge] (d) .. controls (0.5, 1.2) .. (c);
      \end{tikzpicture}
      \caption{A graph $G = (V,E)$ and a subset~$E'$ of edges, in dashed red.}
    \end{subfigure}
    \begin{subfigure}[t]{0.3\textwidth}
      \centering
      \begin{tikzpicture}[scale=1.5]
        \vertex (a) at (0,0) {};
        \vertex (b) at (0,1) {};
        \vertex (e) at (1,0) {};
        \vertex (f) at (1,1) {};
        
        \path[]
        (e) edge[edge] (f)
        ;
        \draw[edge] (a) .. controls (-0.2, 0.5) .. (b);
        \draw[edge] (a) .. controls (0.2, 0.5) .. (b);
        \draw[edge] (b) .. controls (0.4, 1.4) and (-0.4, 1.4) .. (b);
      \end{tikzpicture}
      \caption{The graph $G \quotient E'$.}
    \end{subfigure}
    \begin{subfigure}[t]{0.3\textwidth}
      \centering
      \begin{tikzpicture}[scale=1.5]
        \vertex (a) at (0,0) {};
        \vertex (b) at (1,0) {};
        \vertex (c) at (1,1) {};
        \vertex (d) at (0,1) {};
        \vertex (e) at (2,0) {};
        \vertex (f) at (2,1) {};
        
        \path[]
        (d) edge[edge] (c)
        (a) edge[edge] (b)
        (b) edge[edge] (c)
        (e) edge[edge] (f)
        ;
      \end{tikzpicture}
      \caption{The graph $G \subtract E'$.}
    \end{subfigure}
  \end{center}
  \vspace{-8pt}
  \caption{Example of the two constructions in 
    Definition~\ref{definition:quotient_graphs}.}
  \label{figure:quotient_graphs}
\end{figure}

\begin{definition}
  \label{definition:pseudo_laman}
  Let $B = (G, H)$ be a bigraph with biedges $\edg$, then we say that $B$ is
  \emph{pseudo-Laman} if $\dim(G) + \dim(H) \, = \, |\edg| + 1$,
  where $\dim(G) \, := \, |V| - |\{ \text{connected components of } G \}|$.
\end{definition}
It can be easily seen, that if $G$ is a Laman graph, then the bigraph $(G,G)$
is pseudo-Laman.  From a given bigraph we want to construct new ones with a
smaller number of edges.  We introduce two constructions, \emph{quotient} and
\emph{complement}, both for usual graphs (see
Figure~\ref{figure:quotient_graphs}) and for bigraphs.
\begin{definition}
  \label{definition:quotient_graphs}
  Let $G = (V, E) $ be a graph, and let $E' \subseteq E$. We define two new
  graphs, denoted $G \quotient E'$ and $G \subtract E'$, as follows:
  \begin{itemize}
  \item Let $G'$ be the subgraph of $G$ determined by $E'$.  Then we define $G
    \quotient E'$ to be the graph obtained as follows: its vertices are the
    equivalence classes of the vertices of $G$ modulo the relation dictating
    that two vertices $u$ and $v$ are equivalent if there exists a path in
    $G'$ connecting them; its edges are determined by edges in $E \setminus
    E'$.
  \item Let $\hat{V}$ be the set of vertices of $G$ that are endpoints of some
    edge not in $E'$. Set $\hat{E} = E \setminus E'$. Define $G \subtract E' =
    (\hat{V}, \hat{E})$.
  \end{itemize}
\end{definition}

\begin{definition}
  Let $B = \bigl( G, H \bigr)$ be a bigraph, where $G = (V, \edg)$ and $H = (W, \edg)$.
  Given $\mcal{M} \subseteq \edg$, we define two bigraphs 
  $\leftquot{\mcal{M}}{B}  = \bigl( G \quotient \mcal{M}, \, H \subtract \mcal{M} \bigr)$ and 
  $\rightquot{\mcal{M}}{B} = \bigl( G \subtract \mcal{M}, \, H \quotient \mcal{M} \bigr)$,
  with the same set of biedges $\edg' = \edg \setminus \mcal{M}$.
\end{definition}

\begin{theorem}
  \label{theorem:laman_number}
  Let $B=(G,H)$ be a pseudo-Laman bigraph with biedges $\edg$. Let $\bar{e}\in\edg$
  be a fixed biedge, then
  \begin{itemize}
  \item If $G$ or $H$ has a self-loop, then $\Lam(B) = 0$.
  \item If both $G$ and $H$ consist of a single edge joining two different
    vertices, then $\Lam(B) = 1$.
  \item Otherwise 
    \[
      \Lam(B) = 
      \Lam \bigl( {}^{\{ \bar{e} \} } \! B \bigr) +  
      \Lam \bigl( B^{\{ \bar{e} \}} \bigr) + 
      \sum_{(\mcal{M}, \mcal{N})} 
      \Lam \bigl( \leftquot{\mcal{M}}{B} \bigr) \cdot 
      \Lam \bigl( \rightquot{\mcal{N}}{B} \bigr),
    \]
    where each pair $(\mcal{M}, \mcal{N}) \subseteq \edg^2$ satisfies
    $\mcal{M} \cup \mcal{N} = \edg$ and $\mcal{M} \cap \mcal{N} = \{ \bar{e} \}$,
    and where $\mcal{M}$ and $\mcal{N}$ are such that $|\mcal{M}| \geq 2$, $|\mcal{N}| \geq 2$,
    and both $\leftquot{\mcal{M}}{B}$ and $\rightquot{\mcal{N}}{B}$ are pseudo-Laman.
  \end{itemize}
\end{theorem}

Although the algorithm resulting from Theorem~\ref{theorem:laman_number} has
exponential complexity, it is much faster than computing the Laman number from
a parametrized system of polynomial equations, even if the parameters are
substituted by random values. Using our recursion formula we were able to
compute Laman numbers for all Laman graphs up to 13 vertices.  Furthermore, we
computed Laman numbers for single graphs up to 22 vertices, which was out of
reach with the previous methods.  Additional information including
implementations in Mathematica and in \verb!C++! can be found at
\url{www.koutschan.de/data/laman/} and \url{https://zenodo.org/record/1245506}.

\section{Bounds on the Number of Realizations}

The first \emph{upper bound}~\cite{Borcea2004} on the number of realizations
of rigid graphs was derived using degree bounds from algebraic geometry. Based
on the theory of distance matrices and determinantal varieties, the upper
bound $\binom{2n-4}{n-2}=\Theta( 4^{n}/ \sqrt{n})$ is obtained, where $n$
denotes the number of vertices.  This bound was improved~\cite{Steffens2010}
by exploiting the sparsity of the underlying polynomial systems, and it was
further improved by applying additional tricks to take advantage of the
sparsity and the common sub-expressions that appear in the polynomial
systems~\cite{Emiris2013}.  A direct application of mixed volume techniques,
which capture the sparsity of a polynomial system, yields a bound of
$4^{n-2}$.  If one also takes into account the degree of the vertices, then
for a Laman graph with $k \geq 4$ degree-$2$ vertices, the number of
realizations of~$G$ is bounded from above by $2^{k-4} 4^{n-k}$.

The first \emph{lower bounds} for the number of realizations of Laman graphs
were $24^{\lfloor (n-2)/4 \rfloor}$ (approx.\ $2.21^n$) and $2\cdot
12^{\lfloor (n-3)/3 \rfloor}$ (approx.\ $2.29^n$), which exploited a gluing
process using a caterpillar, resp.\ fan construction~\cite{Borcea2004}.  Both
constructions use the three-prism graph (sometimes also called Desargues
graph) as a building block, which is a graph with $n=6$ vertices and $24$
realizations.  More recent lower bounds are $2.30^n$ from~\cite{EmirisMoroz}
and $2.41^n$ from~\cite{JacksonOwen2012}.

We derive better lower bounds on the maximal number of complex realizations of
minimally rigid graphs with a prescribed number of vertices.  Clearly, the
number of complex realizations is an upper bound on the number of real
realizations. It is known~\cite{JacksonOwen2012} that the numbers of real and
complex realizations do not match in general, and it is an interesting problem
to quantify this gap. On the one hand, one can construct infinite families of
graphs for which the ratio between real and complex realizations tends to
zero. On the other hand, there are graphs, see~\cite{EmirisMoroz} for a
nontrivial example, where real edge lengths can be found such that there exist
as many real realizations as complex ones.

Using the novel algorithm presented in Section~\ref{sec:laman_number} we
compute the exact number of realizations for graphs with a relatively small
number of vertices.  Then we introduce techniques to ``glue'' an arbitrary
number of such small graphs in order to produce graphs with a high number of
vertices (and edges) that preserve rigidity.  The gluing process allows us to
derive the number of realizations of the final graph from the number of
realizations of its components, and in this way we derive a lower bound for
the number of realizations in~$\C^2$. Moreover, we perform extensive
experiments in order to identify those small graphs that attain the maximum
number of realizations and that can be the building blocks for the gluing
process.

\begin{definition}\label{def:laman-number-2d}
  We define $\maxlamIIn$ to be the largest Laman number that is achieved among
  all Laman graphs with $n$ vertices.
\end{definition}

\subsection{Constructions}
\label{sec:constr-2d}

We discuss different constructions of infinite families of Laman
graphs $(G_n)_{n\in\N}$ with $G_n$ having $n$ vertices. We do this in a way
such that we know precisely the Laman number for each member of the family.
This directly leads to a lower bound on $\maxlamIIn$.
The ideas of these constructions are described in~\cite{Borcea2004};
they were used to get lower bounds by connecting several three-prism graphs at a common basis.
Here, we generalize them in order to connect any Laman graphs at an arbitrary Laman base.
We present three such constructions.

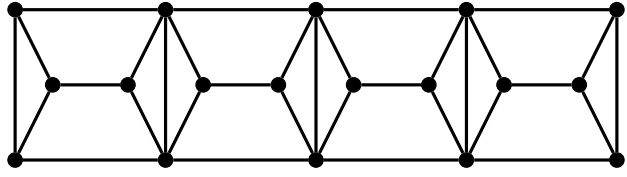
\begin{figure}
  \begin{center}
    \begin{tikzpicture}[scale=1]
      \vertex (a) at (0.00,0.00) {};
      \vertex (b) at (0.00,2.00) {};
      \vertex (c) at (0.50,1.00) {};
      \vertex (d) at (1.50,1.00) {};
      \vertex (e) at (2.00,0.00) {};
      \vertex (f) at (2.00,2.00) {};
      \vertex (g) at (2.50,1.00) {};
      \vertex (h) at (3.50,1.00) {};
      \vertex (i) at (4.00,0.00) {};
      \vertex (j) at (4.00,2.00) {};
      \vertex (k) at (4.50,1.00) {};
      \vertex (l) at (5.50,1.00) {};
      \vertex (m) at (6.00,0.00) {};
      \vertex (n) at (6.00,2.00) {};
      \vertex (o) at (6.50,1.00) {};
      \vertex (p) at (7.50,1.00) {};
      \vertex (q) at (8.00,0.00) {};
      \vertex (r) at (8.00,2.00) {};
      \draw[edge] (a)edge(b) (a)edge(c) (b)edge(c) (a)edge(e) (b)edge(f)
      (c)edge(d) (d)edge(e) (e)edge(f) (d)edge(f) (e)edge(g) (f)edge(g)
      (e)edge(i) (f)edge(j) (g)edge(h) (h)edge(i) (i)edge(j) (h)edge(j)
      (i)edge(k) (j)edge(k) (i)edge(m) (j)edge(n) (k)edge(l) (l)edge(m)
      (m)edge(n) (l)edge(n) (m)edge(o) (n)edge(o) (m)edge(q) (n)edge(r)
      (o)edge(p) (p)edge(q) (q)edge(r) (p)edge(r);
    \end{tikzpicture}
  \end{center}
  \caption{Caterpillar construction with $4$ copies of the three-prism graph.}
  \label{figure:caterpillar}
\end{figure}

The \emph{caterpillar construction}~\cite{Borcea2004} works as follows: place
$k$ copies of a Laman graph $G=(V,E)$ in a row and connect every two
neighboring ones by means of a shared edge (see
Figure~\ref{figure:caterpillar}).  Alternatively, one can let all $k$ graphs
share the same edge. In any case, the resulting assembly has $2+k(|V|-2)$
vertices and its Laman number is $\lamII(G)^k$, since each of the $k$ copies
of $G$ can achieve all its $\lamII(G)$ different realizations, independently
of what happens with the other copies. Hence, among all Laman graphs with
$n=2+k(|V|-2)$ vertices there exists one with $\lamII(G)^k$ realizations. If
the number of vertices $n$ is not of the form $2+k(|V|-2)$ then we can use the
previous caterpillar graph with $\lfloor(n-2)/(|V|-2)\rfloor$ copies of $G$
and perform some Henneberg steps of type~1: such a step adds one vertex and
connects it to two existing vertices, thereby doubling the Laman number.
Summarizing, for any Laman graph~$G$, we obtain the following lower bound from
the caterpillar construction:
\[
  \maxlamIIn \geq 2^{(n-2)\modop(|V|-2)} \cdot \lamII(G)^{\lfloor(n-2)/(|V|-2)\rfloor}\qquad (n\geq2).
\]

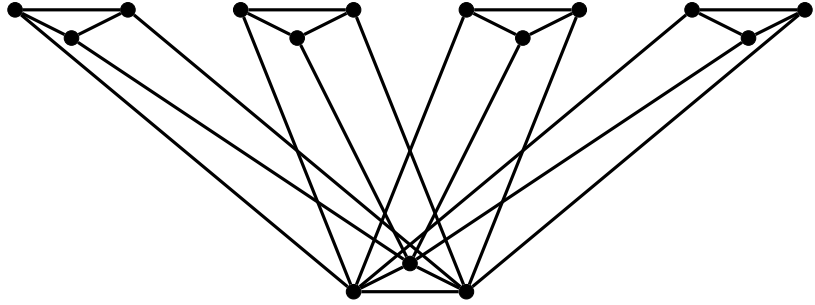
\begin{figure}
  \begin{center}
    \begin{tikzpicture}[scale=0.75]
      \vertex (a) at (-1.00,0.00) {};
      \vertex (b) at (1.00,0.00) {};
      \vertex (c) at (0.00,0.50) {};
      \vertex (d) at (-7.00,5.00) {};
      \vertex (e) at (-5.00,5.00) {};
      \vertex (f) at (-6.00,4.50) {};
      \vertex (g) at (-3.00,5.00) {};
      \vertex (h) at (-1.00,5.00) {};
      \vertex (i) at (-2.00,4.50) {};
      \vertex (j) at (1.00,5.00) {};
      \vertex (k) at (3.00,5.00) {};
      \vertex (l) at (2.00,4.50) {};
      \vertex (m) at (5.00,5.00) {};
      \vertex (n) at (7.00,5.00) {};
      \vertex (o) at (6.00,4.50) {};
      \draw[edge] (a)edge(b) (b)edge(c) (c)edge(a) (d)edge(e) (e)edge(f)
      (f)edge(d) (d)edge(a) (e)edge(b) (f)edge(c) (g)edge(h) (h)edge(i)
      (i)edge(g) (g)edge(a) (h)edge(b) (i)edge(c) (j)edge(k) (k)edge(l)
      (l)edge(j) (j)edge(a) (k)edge(b) (l)edge(c) (m)edge(n) (n)edge(o)
      (o)edge(m) (m)edge(a) (n)edge(b) (o)edge(c);
    \end{tikzpicture}
  \end{center}
  \caption{Fan construction with $4$ copies of the three-prism graph.}
  \label{figure:fan}
\end{figure}

The second construction is the \emph{fan construction}: take a
Laman graph $G=(V,E)$ that contains a triangle (i.e., a $3$-cycle), and glue $k$ copies of $G$
along that triangle (see Figure~\ref{figure:fan}). Once we fix one of the two
possible realizations of that triangle, each copy of $G$ admits $\lamII(G)/2$
realizations. The remaining $\lamII(G)/2$ realizations are obtained by mirroring, i.e.,
by using the second realization of the common triangle. Similarly as before, the
assembled fan is a Laman graph with $3+k(|V|-3)$ vertices that admits
$2\cdot(\lamII(G)/2)^k$ realizations. Hence, we get the following lower bound:
\[
  \maxlamIIn \geq 2^{(n-3)\modop(|V|-3)}\cdot 2\cdot\left(\frac{\lamII(G)}{2}\right)^{\!\lfloor(n-3)/(|V|-3)\rfloor} \qquad (n\geq3).
\]
While the caterpillar construction can be done with any Laman graph, this is
not the case with the fan. For example, the Laman graph with 12 vertices displayed in 
Figure~\ref{figure:max_graphs} has no $3$-cycle and therefore cannot be used for
the fan construction.

\begin{figure}
  \begin{center}
    \setlength{\tabcolsep}{18pt}
    \begin{tabular}{cccc}
      \begin{tikzpicture}
        \vertex (a) at (0,0) {};
	\vertex (b) at (1,0) {};
	\vertex (c) at (0.5,0.866025) {};
	\draw[edge] (a)edge(b) (b)edge(c) (a)edge(c);
      \end{tikzpicture}
      &
      \begin{tikzpicture}
        \vertex (a) at (0,0) {};
	\vertex (b) at (1,0) {};
	\vertex (c) at (0.5,0.866025) {};
	\vertex (d) at (1.5,0.866025) {};
	\draw[edge] (a)edge(b) (b)edge(c) (a)edge(c) (b)edge(d) (c)edge(d);
      \end{tikzpicture}
      &
      \begin{tikzpicture}[rotate=-60]
        \vertex (a) at (0,0) {};
	\vertex (b) at (1,0) {};
	\vertex (c) at (0.5,0.866025) {};
	\vertex (d) at (1.5,0.866025) {};
	\vertex (e) at (0,1.732050) {};
	\draw[edge] (a)edge(b) (b)edge(c) (a)edge(c) (b)edge(d) (c)edge(d) (a)edge(e) (d)edge(e);
      \end{tikzpicture}
      &
      \begin{tikzpicture}[yscale=0.04,xscale=0.05,rotate=90]
        \vertex (a) at (-15.,-18.4482) {};
	\vertex (b) at (15,-18.4482) {};
	\vertex (c) at (0.,-6.78132) {};
	\vertex (d) at (-15.,18.4482) {};
	\vertex (e) at (15.,18.4482) {};
	\vertex (f) at (0.,6.78132) {};
	\draw[edge]  (a)edge(b) (a)edge(c) (a)edge(d) (b)edge(c) (b)edge(e) (c)edge(f) (d)edge(e) (d)edge(f) (e)edge(f);
      \end{tikzpicture}
      \\[1ex]
      triangle & $H_1$ & $H_2$ & $H_3$ ``three-prism''
    \end{tabular}
  \end{center}
  \caption{Bases for the generalized fan construction and their encodings.}
  \label{figure:fanbases}
\end{figure}
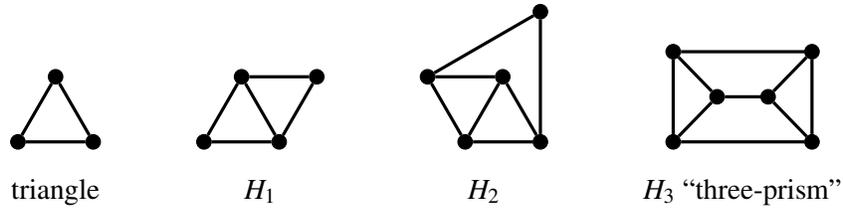

As a third construction, we propose the \emph{generalized fan construction}:
instead of a triangle, we may use any Laman subgraph $H=(W,F)$ of $G$ for
gluing. Using $k$ copies of $G$, we end up with a fan consisting of
$|W|+k(|V|-|W|)$ vertices and Laman number at least $\lamII(H)\cdot(\lamII(G)/\lamII(H))^k$. Here we
assume that the realizations of $G$ are divided into $L(H)$ equivalence classes
of equal size, by considering two realizations of $G$ as equivalent if the
induced realizations of $H$ are equal (up to rotations and translations). If
this assumption was violated, the resulting lower bound would be even better;
thus we can safely state the following bound:
\[
  \maxlamIIn \geq 2^{(n-|W|)\modop(|V|-|W|)} \cdot\lamII(H) \cdot
  \left(\frac{\lamII(G)}{\lamII(H)}\right)^{\!\lfloor(n-|W|)/(|V|-|W|)\rfloor} \qquad (n\geq|W|).
\]
Note that the previously described fan construction is a special instance of the
generalized fan, by taking as the subgraph $H$ a triangle with $\lamII(H)=2$.
To indicate the subgraph of a generalized fan construction we also write $H$-fan.
The fan fixing the 4-vertex Laman graph is then denoted by $H_1$-fan (see
Figure~\ref{figure:fanbases} for these base graphs and their naming convention).

\subsection{Lower bounds}\label{sec:graphs-2d}

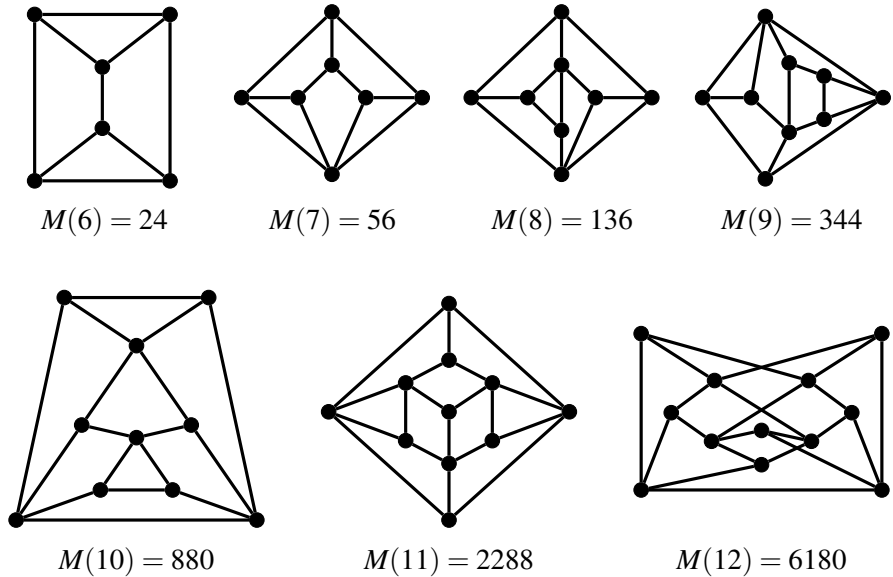
\begin{figure}
  \begin{center}
    \begin{tabular}{c@{\hspace{12pt}}c@{\hspace{12pt}}c@{\hspace{12pt}}c}
      \begin{tikzpicture}[scale=0.06] 
	\draw[white] (-21,22) rectangle (22,-21);
	\vertex (a) at (-15.,-18.4482) {};
	\vertex (b) at (15,-18.4482) {};
	\vertex (c) at (0.,-6.78132) {};
	\vertex (d) at (-15.,18.4482) {};
	\vertex (e) at (15.,18.4482) {};
	\vertex (f) at (0.,6.78132) {};
	\draw[edge]  (a)edge(b) (a)edge(c) (a)edge(d) (b)edge(c) (b)edge(e) (c)edge(f) (d)edge(e) (d)edge(f) (e)edge(f);
      \end{tikzpicture}
      &
      \begin{tikzpicture}[scale=0.06] 
	\draw[white] (-21,22) rectangle (22,-21);
	\vertex (a) at (0.00,-17.00) {};
	\vertex (b) at (20.00,0) {};
	\vertex (c) at (7.50,0) {};
	\vertex (d) at (-20.00,0) {};
	\vertex (e) at (-7.50,0) {};
	\vertex (f) at (0.00,19.00) {};
	\vertex (g) at (0.00,7.24) {};
	\draw[edge]  (a)edge(b) (a)edge(c) (a)edge(d) (a)edge(e) (b)edge(c)
	(b)edge(f) (c)edge(g) (d)edge(e) (d)edge(f) (e)edge(g) (f)edge(g);
      \end{tikzpicture}
      &
      \begin{tikzpicture}[scale=0.06] 
	\draw[white] (-21,22) rectangle (22,-21);
	\vertex (a) at (-0.00,-17.00) {};
	\vertex (b) at (-0.00,7.24) {};
	\vertex (c) at (7.50,0.00) {};
	\vertex (d) at (-0.00,-7.24) {};
	\vertex (e) at (20.00,0.00) {};
	\vertex (f) at (-20.00,0.00) {};
	\vertex (g) at (-7.50,0.00) {};
	\vertex (h) at (-0.00,19.00) {};
	\draw[edge]  (a)edge(c) (a)edge(d) (a)edge(e) (a)edge(f) (b)edge(c)
	(b)edge(d) (b)edge(g) (b)edge(h) (c)edge(e) (d)edge(g) (e)edge(h)
	(f)edge(g) (f)edge(h);
      \end{tikzpicture}
      &
      \begin{tikzpicture}[scale=0.06] 
	\draw[white] (-21,22) rectangle (22,-21);
	\vertex (a) at (21.00,0.00) {};
	\vertex (b) at (-5.10,18.00) {};
	\vertex (c) at (0.17,-7.70) {};
	\vertex (d) at (7.90,-4.80) {};
	\vertex (e) at (-5.10,-18.00) {};
	\vertex (f) at (0.17,7.70) {};
	\vertex (g) at (-8.20,0.00) {};
	\vertex (h) at (7.90,4.80) {};
	\vertex (i) at (-19.00,0.00) {};
	\draw[edge]  (a)edge(b) (a)edge(d) (a)edge(e) (a)edge(h) (b)edge(f)
	(b)edge(g) (b)edge(i) (c)edge(d) (c)edge(e) (c)edge(f) (c)edge(g)
	(d)edge(h) (e)edge(i) (f)edge(h) (g)edge(i);
      \end{tikzpicture}
      \\
      $\Mtwo(6)=24$ & $\Mtwo(7)=56$ & $\Mtwo(8)=136$ & $\Mtwo(9)=344$
    \end{tabular}\\[18pt]
    \begin{tabular}{c@{\hspace{4pt}}c@{\hspace{4pt}}c}
      \begin{tikzpicture}[scale=0.08] 
	\draw[white] (-25,22) rectangle (25,-18);
	\vertex (a) at (-20.00,-16.00) {};
	\vertex (b) at (20.00,-16.00) {};
	\vertex (c) at (-0.00,-2.30) {};
	\vertex (d) at (-0.00,13.00) {};
	\vertex (e) at (-9.10,-0.20) {};
	\vertex (f) at (9.10,-0.20) {};
	\vertex (g) at (-6.00,-11.00) {};
	\vertex (h) at (-12.00,21.00) {};
	\vertex (i) at (6.00,-11.00) {};
	\vertex (j) at (12.00,21.00) {};
	\draw[edge]  (a)edge(b) (a)edge(e) (a)edge(g) (a)edge(h) (b)edge(f)
	(b)edge(i) (b)edge(j) (c)edge(e) (c)edge(f) (c)edge(g) (c)edge(i)
	(d)edge(e) (d)edge(f) (d)edge(h) (d)edge(j) (g)edge(i) (h)edge(j);
      \end{tikzpicture}
      &
      \begin{tikzpicture}[scale=0.08] 
	\draw[white] (-25,20) rectangle (25,-20);
	\vertex (a) at (7.20,4.80) {};
	\vertex (b) at (20.00,-0.00) {};
	\vertex (c) at (-7.20,4.80) {};
	\vertex (d) at (-20.00,-0.00) {};
	\vertex (e) at (0.00,-8.60) {};
	\vertex (f) at (7.20,-4.80) {};
	\vertex (g) at (0.00,-0.00) {};
	\vertex (h) at (0.00,-18.00) {};
	\vertex (i) at (-7.20,-4.80) {};
	\vertex (j) at (0.00,8.60) {};
	\vertex (k) at (0.00,18.00) {};
	\draw[edge]  (a)edge(b) (a)edge(f) (a)edge(g) (a)edge(j) (b)edge(f)
	(b)edge(h) (b)edge(k) (c)edge(d) (c)edge(g) (c)edge(i) (c)edge(j)
	(d)edge(h) (d)edge(i) (d)edge(k) (e)edge(f) (e)edge(g) (e)edge(h)
	(e)edge(i) (j)edge(k);
      \end{tikzpicture}
      &
      \begin{tikzpicture}[scale=0.08] 
	\draw[white] (-25,21) rectangle (25,-19);
	\vertex (a) at (20.00,-12.00) {};
	\vertex (b) at (-20.00,-12.00) {};
	\vertex (c) at (7.80,6.20) {};
	\vertex (d) at (-8.30,-3.90) {};
	\vertex (e) at (8.30,-3.90) {};
	\vertex (f) at (-7.80,6.20) {};
	\vertex (g) at (15.00,0.85) {};
	\vertex (h) at (20.00,14.00) {};
	\vertex (i) at (0.00,-2.10) {};
	\vertex (j) at (-20.00,14.00) {};
	\vertex (k) at (0.00,-7.80) {};
	\vertex (l) at (-15.00,0.85) {};
	\draw[edge] (a)edge(b) (a)edge(g) (a)edge(h) (a)edge(i) (b)edge(j)
	(b)edge(k) (b)edge(l) (c)edge(d) (c)edge(g) (c)edge(h) (c)edge(j)
	(d)edge(i) (d)edge(k) (d)edge(l) (e)edge(f) (e)edge(g) (e)edge(i)
	(e)edge(k) (f)edge(h) (f)edge(j) (f)edge(l);
      \end{tikzpicture}
      \\
      $\Mtwo(10)=880$ & $\Mtwo(11)=2288$ & $\Mtwo(12)=6180$
    \end{tabular}
  \end{center}\vspace{-8pt}
  \caption{Unique Laman graphs with $6\leq n\leq12$ with maximal number of realizations}
  \label{figure:max_graphs}
\end{figure}

In order to get good lower bounds, we need particular Laman graphs
that have a large number of realizations. For this purpose we have computed the Laman numbers
of all Laman graphs with up to $n=13$ vertices. 
For each $3 \leq n \leq 12$ we have
identified the (unique) Laman graph with the highest number of
realizations. We present these numbers and the corresponding graphs for
$6\leq n\leq12$ in Figure~\ref{figure:max_graphs}.

However, there are 44\,176\,717 Laman graphs with 12 vertices, and it took 56
processor days to compute the Laman numbers of all of them. Going through all
1\,092\,493\,042 Laman graphs with 13 vertices was an even more challenging
undertaking, and it is unrealistic to do the same
for larger Laman graphs. In order to proceed further, we
developed some heuristics to construct graphs with very high Laman numbers,
albeit not necessarily the highest one. 

\begin{table}
  \begin{center}
    \begin{tabular}{rllllll}
      $n$ & caterpillar & fan        &      $H_1$-fan & $H_2$-fan & $H_3$-fan\\
      \hline\rule{0pt}{14pt}
      6   & 2.21336     & 2.28943    &          2       & 2       & -       \\
      7   & 2.23685     & 2.30033    &          2.28943 & 2       & 2       \\
      8   & 2.26772     & 2.32542    &          2.30033 & 2.28943 & 2       \\
      9   & 2.30338     & 2.35824    &          2.35216 & 2.30033 & 2.28943 \\
      10  & 2.33378     & 2.38581    &          2.35824 & 2.35216 & 2.30033 \\
      11  & 2.36196     & 2.41159    &          2.38581 & 2.35824 & 2.35216 \\
      12  & 2.39386     & 2.43198    &          2.43006 & 2.39802 & 2.35824 \\
      13  & 2.40453     & 2.44498    &          2.44772 & 2.42197 & 2.39802 \\
      14  & 2.43185     & 2.46087    &          2.46391 & 2.44251 & 2.42197 \\
      15  & 2.44695     & 2.47445    &          2.47076 & 2.45031 & 2.42906 \\
      16  & 2.46890     & 2.48657    &          2.48794 & 2.47166 & 2.43712 \\
      17  & 2.48875     & 2.49779    &          2.49160 & 2.48043 & 2.46341 \\
      18  & 2.49378     & 2.50798    & \\
    \end{tabular}
  \end{center}
  \caption{Growth rates (rounded) of the lower bounds.
    For $n\leq13$ these values
    are proven to be the best achievable ones; for $n>13$ the values are
    just the best we found by experiments, hence it is possible that
    there are better ones.
    The drawings of the graphs corresponding to
    the last three columns are given in Figure~\ref{figure:fanbases}.
  }
  \label{table:bounds}
\end{table}

We now use these results to derive new and better lower bounds than the
previously known ones. We apply the caterpillar construction to the Laman
graphs with the maximal number of realizations for $6\leq n\leq13$, and for
$14\leq n\leq18$ we use graphs with high Laman numbers that were found
heuristically.  The fan construction is applied to the maximal Laman graphs
for $6\leq n\leq11$ only, since it is not applicable to the maximal graph with
$12$ vertices, because that graph does not contain $K_3$ as a subgraph (see
Figure~\ref{figure:max_graphs}). All lower bounds that we obtained by these
constructions are summarized in Table~\ref{table:bounds}.

\providecommand{\bibitemdeclare}[2]{}
\providecommand{\url}[1]{\texttt{#1}}
\providecommand{\href}[2]{\texttt{#2}}
\providecommand{\urlalt}[2]{\href{#1}{#2}}
\providecommand{\doi}[1]{doi:\urlalt{http://dx.doi.org/#1}{#1}}
\providecommand{\eprint}[1]{arXiv:\urlalt{https://arxiv.org/abs/#1}{#1}}
\providecommand{\bibinfo}[2]{#2}

\bibliographystyle{eptcs}

\end{document}